\documentclass[epj]{webofc}
\usepackage[utf8]{inputenc}
\usepackage[varg]{txfonts}   
\usepackage{booktabs}
\usepackage{xcolor}
\definecolor{darkred}{rgb}{0.4,0.0,0.0}
\definecolor{darkgreen}{rgb}{0.0,0.4,0.0}
\definecolor{darkblue}{rgb}{0.0,0.0,0.4}
\usepackage[bookmarks,linktocpage,colorlinks,
    linkcolor = darkred,
    urlcolor  = darkblue,
    citecolor = darkgreen]{hyperref}
\newcommand{\bea}{\begin{eqnarray}}
\newcommand{\eea}{\end{eqnarray}} 
\newcommand{\beq}{\begin{equation}}
\newcommand{\eeq}{\end{equation}}

\renewcommand{\b}{\beta}
\renewcommand{\l}{\lambda}

\newcommand{\E}{\mathcal{E}}

\newcommand{\tr}{\text{Tr}}

\newcommand{\bx}{\mathbf{x}}
\newcommand{\by}{\mathbf{y}}
\newcommand{\bk}{\mathbf{k}}

\newcommand{\qbar}{\overline{q}}

\newcommand{\vx}{\bx}
\newcommand{\vy}{\by}
\newcommand{\vz}{\mathbf{z}}
\newcommand{\vk}{\bk}

\newcommand{\vR}{\mathbf{R}}

\newcommand{\m}{\mu}
\newcommand{\g}{\gamma}

\newcommand{\s}{\sigma}

\newcommand{\G}{{\cal G}}

\newcommand{\vp}{\vec{p}}
\newcommand{\N}{{\cal N}}

\newcommand{\vph}{\varphi}
\newcommand{\oh}{\frac{1}{2}}

\newcommand{\dg}{\dagger}
\newcommand{\non}{\nonumber}

\newcommand{\rf}[1]{(\ref{#1})}
\newcommand{\ra}{\rightarrow}
\newcommand{\pa}{\partial}
\usepackage{subfigure}
\wocname{EPJ Web of Conferences}
\woctitle{Lattice2017}
\begin{document}
%
\selectlanguage{english}
\title{%
Observation of a Coulomb flux tube
}
\author{%
\firstname{Jeff} \lastname{Greensite}\inst{1}\fnsep\thanks{Speaker. \email{greensit@sfsu.edu}.  Work supported by the US Dept.\ of Energy under Grant No. DE-SC0013682.} \and
\firstname{Kristian} \lastname{Chung}\inst{1}  }
\institute{%
Physics and Astronomy Dept., San Francisco State University, San Francisco CA 94132 USA
}
\abstract{%
In Coulomb gauge there is a longitudinal color electric field associated with a static quark-antiquark pair. We have measured the spatial distribution of this field, and find that it falls off exponentially with transverse distance from a line joining the two quarks. In other words there is a Coulomb flux tube, with a width that is somewhat smaller than that of the minimal energy flux tube associated with the asymptotic string tension. A confinement criterion for gauge theories with matter fields is also proposed.}
\maketitle
\section{Introduction}\label{intro}

In this talk we will briefly discuss two somewhat related topics:  (i) the collimation of the color Coulomb electric field into a flux tube; and
(ii) a confinement criterion for gauge theories with matter fields in the fundamental representation.  The first topic has appeared in \cite{Chung:2017ref}, while the second, which is work by the speaker and K.\ Matsuyama, is reported in much greater detail in \cite{Greensite:2017ajx}.

\section{The color Coulomb potential}\label{sec-1}

The color Coulomb energy ${\cal E}_C(R)$ is the energy above the vacuum energy ${\cal E}_{vac}$ of the state $\Psi_{\qbar q}$, which is generated by Coulomb gauge quark-antiquark creation operators acting on the ground state $\Psi_0$, i.e.
\beq
             {\cal E}_C(R)  = \langle \Psi_{\qbar q}|H|\Psi_{\qbar q} \rangle -  {\cal E}_{vac} \ ,
\eeq
where, for heavy quark-antiquarks separated by $R=|\vR_1-\vR_2|$ (and $\s$ a color index)
\beq
           |\Psi_{\qbar q}\rangle =  \N \int {d^3k_1 \over (2\pi)^3}  {d^3k_2 \over (2\pi)^3} b^{\dg \s}(k_1,\lambda_1) 
           d^{\dg \s}(k_2,\lambda_2)   e^{-i(\vk_1 \cdot \vR_1 + \vk_2 \cdot \vR_2)}  |\Psi_0 \rangle \ .
\eeq
The interaction energy is due to the fact that, in Coulomb gauge, creation of a charged color source is automatically accompanied
by a longitudinal color electric field, due to the Gauss law constraint $D_i E_i = \rho_q$.   In Coulomb gauge it works this way: separate the E-field into a transverse and longitudinal part  $E=E^{tr} + E_L,~~~ E_L = -\nabla \phi $, then $-\pa_i D_i \phi = \rho_q + \rho_g$,
where $\rho^a_q = g\qbar T^a \g_0 q$ and  $\rho^a_g = g f^{abc} E_k^{tr,b} A_k^c$.  Defining the ghost operator 
\beq 
            G^{ab}(\vx,\vy;A)  =   \left({1 \over - \pa_i D_i(A)}\right)^{ab}_{\vx \vy} \ ,
\eeq
the solution of the Gauss law constraint is
\beq
          \vec{E}^a_L(\vx,A,\rho) =   -\vec{\nabla}_x \int d^3y ~  G^{ab}(\vx,\vy;A) (\rho_q^b(\vy)+\rho^b_g(\vy)) \ ,
\eeq
and in the Hamiltonian $\int E_L^2$ gives rise to the non-local Coulomb interaction operator
\beq
    {\cal E}_{coul} =  \int d^3x d^3y d^3z ~ \rho^a(\vx)   \left( G^{ac}(\vx,\vz,A) (-\nabla^2)_{\vz}  G^{cb}(\vz,\vy,A) \right) \rho^b(\vy) \ .
\label{Ecoul}
\eeq

   We know from computer simulations that the Coulomb energy $ {\cal E}_C(R)$ rises linearly with $R$ 
 \cite{Greensite:2003xf,Greensite:2004ke}.  But what is the
spatial distribution of $E_L^2$ due to the static color charges?   There is no obvious reason that it should be concentrated in a flux tube.  If we consider \emph{only} $E_L$ due to the static quark-antiquark pair we have
\beq
\vec{E}^a_{L,q\qbar}(\vx,A,\rho_q) =   -\vec{\nabla}_x \int d^3y ~  G^{ab}(\vx,\vy;A) \rho_q^b(\vy) \ ,
\eeq
then squaring, summing over the color index, and taking the expectation value of the matter field color charge densities
leads to
\bea
        E^2_{L,q\qbar}(\vx,A)   &=&     {g^2 \over 2N_c} \Bigl(  \nabla_x G^{ab}(\vx,0;A) \cdot \nabla_x G^{ab}(\vx,0;A) 
  +  \nabla_x G^{ab}(\vx,\vR;A) \cdot \nabla_x G^{ab}(\vx,\vR;A) \non \\                    
  & &  \quad        - 2 \nabla_x G^{ab}(\vx,0;A) \cdot \nabla_x G^{ab}(\vx,\vR;A)   \Bigr)      \ .
\eea
It seems unlikely that $G^{ab}(\vx,\vy,A)$ would fall exponentially with $|\vx-\vy|$ for typical vacuum configurations.  In that case it would be hard to see how the Coulomb potential could rise linearly with $R$.  Also the momentum-space ghost propagator $G^{ab}(\vk)$, has been computed in lattice Monte Carlo simulations \cite{Burgio:2012bk,Nakagawa:2009zf,Langfeld:2004qs}  with the position space result $G^{ab}(r) \sim \delta^{ab}/r^{0.56}$ in the infrared.
So it is reasonable to assume some power-law falloff of $G^{ab}(\vx,\vy,A)$ with separation  $|\vx-\vy|$,
for typical vacuum fluctuations $A$.  Then, unless there are very delicate cancellations, one would expect a power law falloff  for $E^2_L(\vx,A)$, as the distance of point $\vx$ from the $\qbar q$ sources increases. This would imply a long-range color Coulomb dipole field in the physical state $\Psi_{\qbar q}$.

\section{Lattice measurements}\label{sec-2}
Let
\beq
              L_t(\vx) \equiv T\exp\left[ig\int_0^t dt' A_4(\vx,t') \right] \ .
\eeq
Then the Coulomb energy is obtained from the logarithmic time derivative 
\beq
  {\cal E}_C(R)  = - \lim_{t\ra 0} {d \over dt} \log  \langle \Psi_{\qbar q}|e^{-Ht}|\Psi_{\qbar q} \rangle  = - \lim_{t\ra 0} {d \over dt} \log  \big\langle \tr[L_t({\bf 0}) L_t^\dg({\bf R})] \big\rangle \ ,
\label{Ecoul}
\eeq
while the minimal energy of static quark-antiquark state is obtained in the opposite limit
\beq
            {\cal E}_{min}(R) = - \lim_{t\ra \infty} {d \over dt} \log  \big\langle \tr[L_t({\bf 0}) L_t^\dg({\bf R})] \big\rangle \ .
\eeq
The lattice version is
\beq
{\cal E}_C(R_L) = -    \log \big\langle {1\over N_c}\tr[U_0({\bf 0},0) U_0^\dg({\bf R}_L,0)] \big\rangle \ ,
\eeq
which we measure, convert to physical units using the lattice spacing $a(\b)$, and fit to
\beq
       {\cal E}^{phys}_C(R) = \sigma_c(\beta) R - {\gamma(\beta) \over R} + {c(\beta) \over a(\beta)}  \ .
\eeq

    It was found in lattice Monte Carlo simulations of SU(3) gauge theory that the Coulomb string tension $\s_C$ is about four times greater than the asymptotic string tension $\s$ \cite{Nakagawa:2006fk,Greensite:2014bua}.  It was also found that an $R$-independent self-energy term $c/a(\beta)$ can be isolated and subtracted, and that there is a $-\gamma/R$ term in the potential with $\gamma \ra \pi/12$ in the continuum limit \cite{Greensite:2014bua}. This looks like a L{\"u}scher term.  We find the same result for $\gamma$ in our SU(2) simulations
at $\b=2.5$.  It is natural to ask whether this result for $\g$ is just a coincidence, or instead indicates some connection to string theory.  The odd fact that $\gamma \approx \pi/12$ motivates us to look at the energy distribution of the Coulomb electric field.

\begin{figure}[htb]
\begin{center}
\includegraphics[scale=0.8]{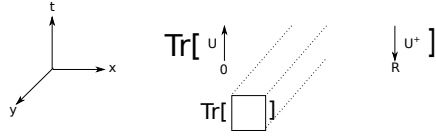}
\end{center}
\caption{The observable for the calculation of the $x$-component of the color electric energy density $Q(R,y)$, generated by a quark-antiquark pair along the $x$-axis separated by distance $R$, as a function of the transverse distance $y$ away from the midpoint. 
$U,U^\dg$ denote timelike link variables at equal times.} 
\label{corr}
\end{figure}

\begin{figure}[t!]
\begin{center}
\subfigure[~$R=2$]  
{   
 \label{r2}
 \includegraphics[scale=0.35]{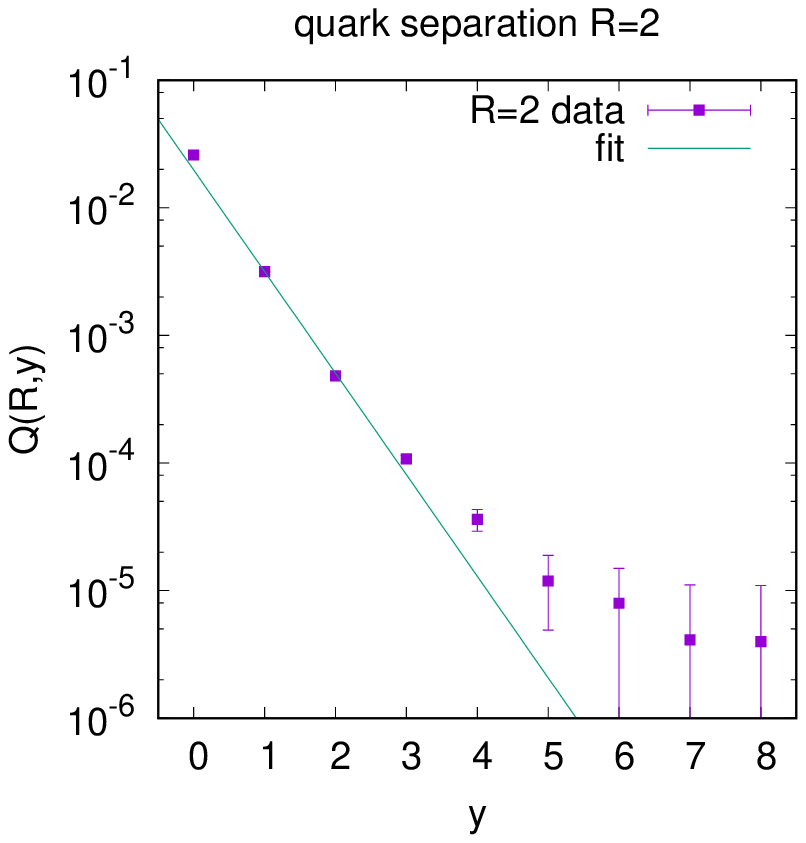}
}
\subfigure[~$R=3$]  
{   
 \label{r3}
 \includegraphics[scale=0.35]{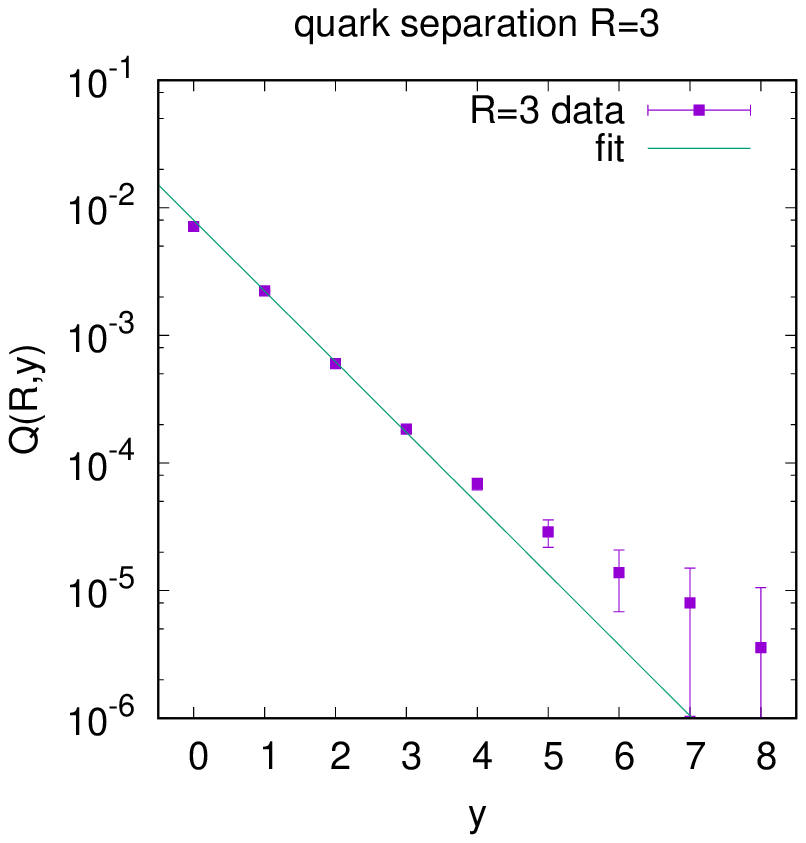}
}
\subfigure[~$R=4$]  
{   
 \label{r4}
 \includegraphics[scale=0.35]{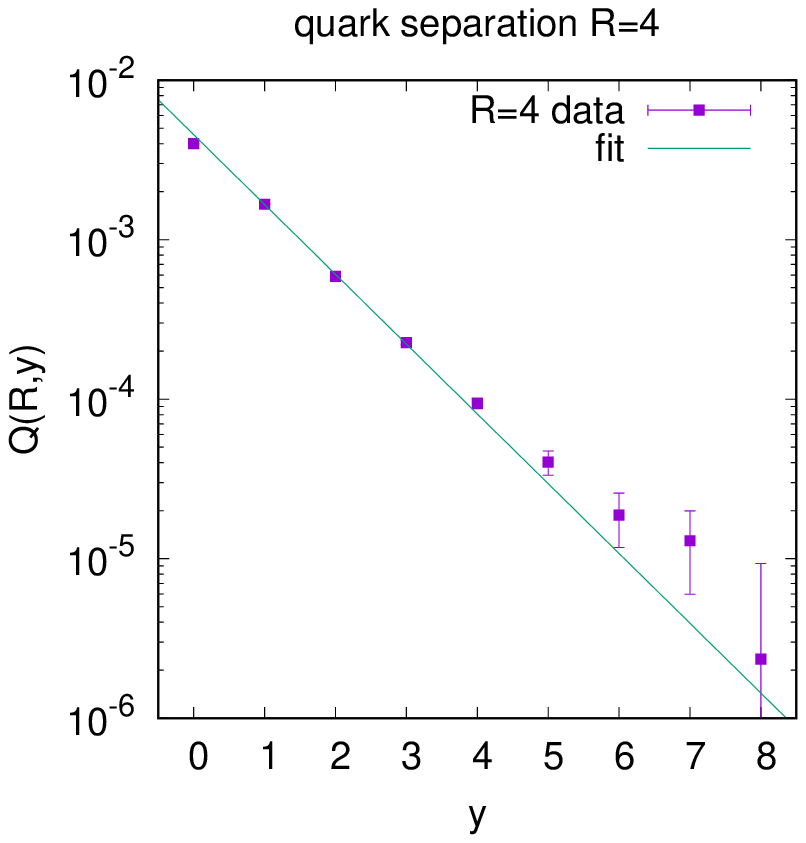}
}
\subfigure[~$R=5$]  
{   
 \label{r5}
 \includegraphics[scale=0.35]{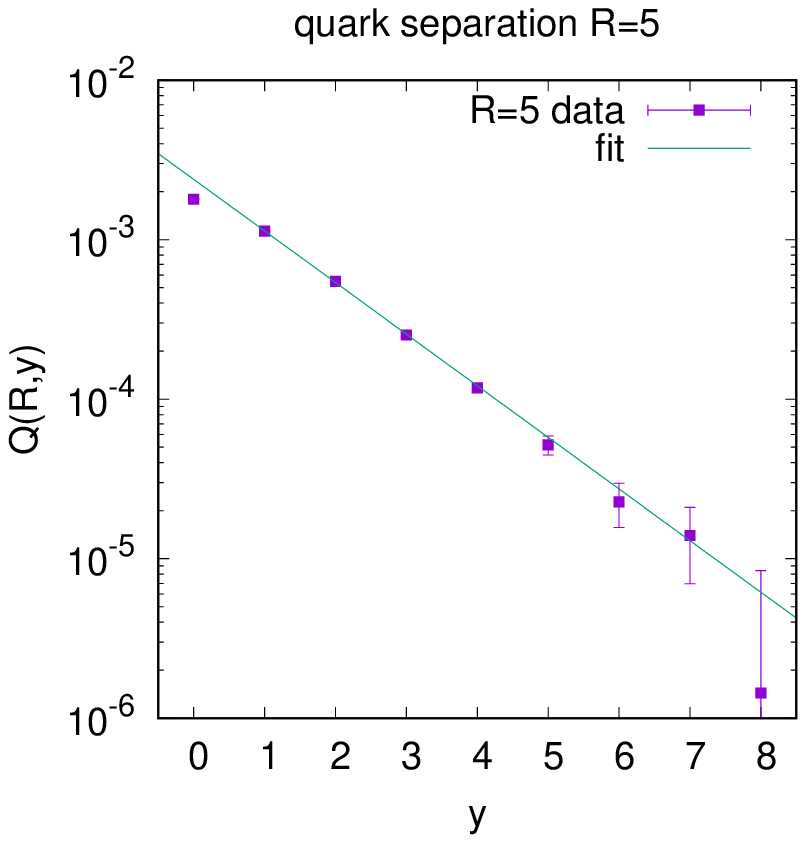}
}
\subfigure[~$R=6$]  
{   
 \label{r6}
 \includegraphics[scale=0.35]{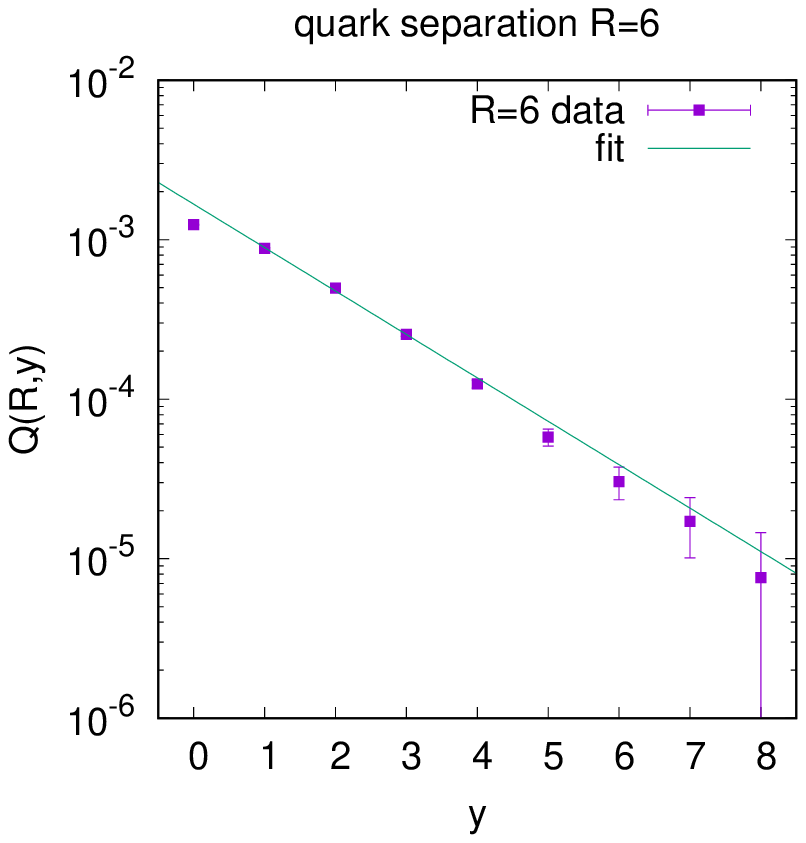}
}
\subfigure[~$R=7$]  
{   
\label{r7}
\includegraphics[scale=0.35]{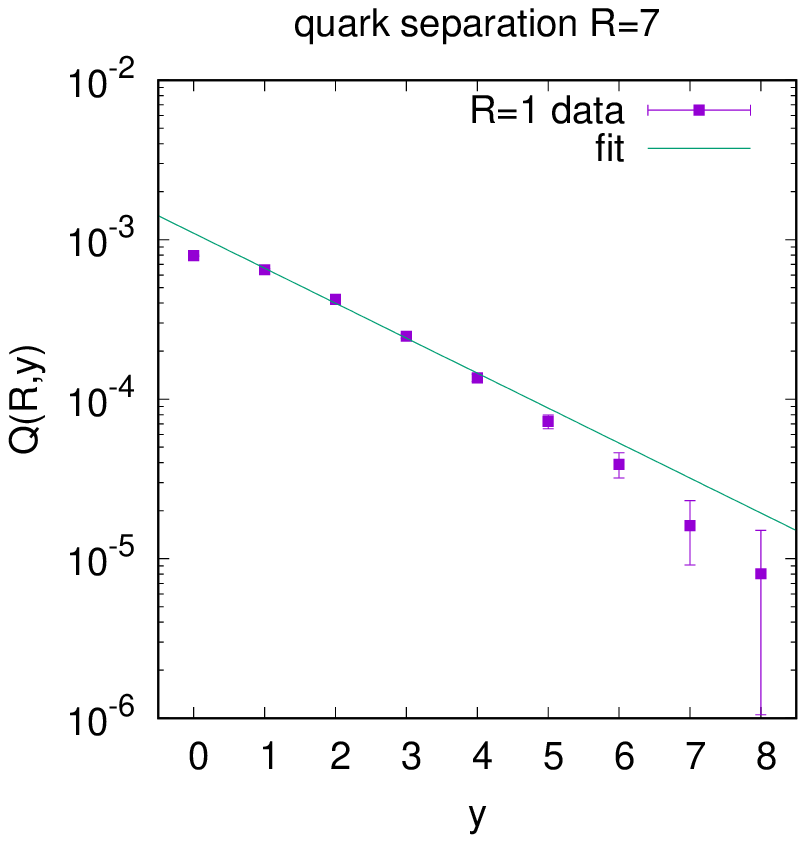}
}
\end{center}
\caption{The connected correlator $Q(R,y)$ of two timelike links and one plaquette, for fixed link separation $R$, vs.\ transverse
separation $y$ of the plaquette from the midpoint of the line of quark-antiquark separation. This is a measure of the
falloff of the color Coulomb energy density with transverse distance away from a quark-antiquark dipole in Coulomb gauge. The simulation is for SU(2) pure gauge theory at $\b=2.5$.  The lines
show a best fit to an exponential falloff $\exp[-(a+by)]$.}
\label{falloff}
\end{figure}

\begin{figure}[h!]
\begin{center}
\subfigure[]  
{   
 \includegraphics[scale=0.135]{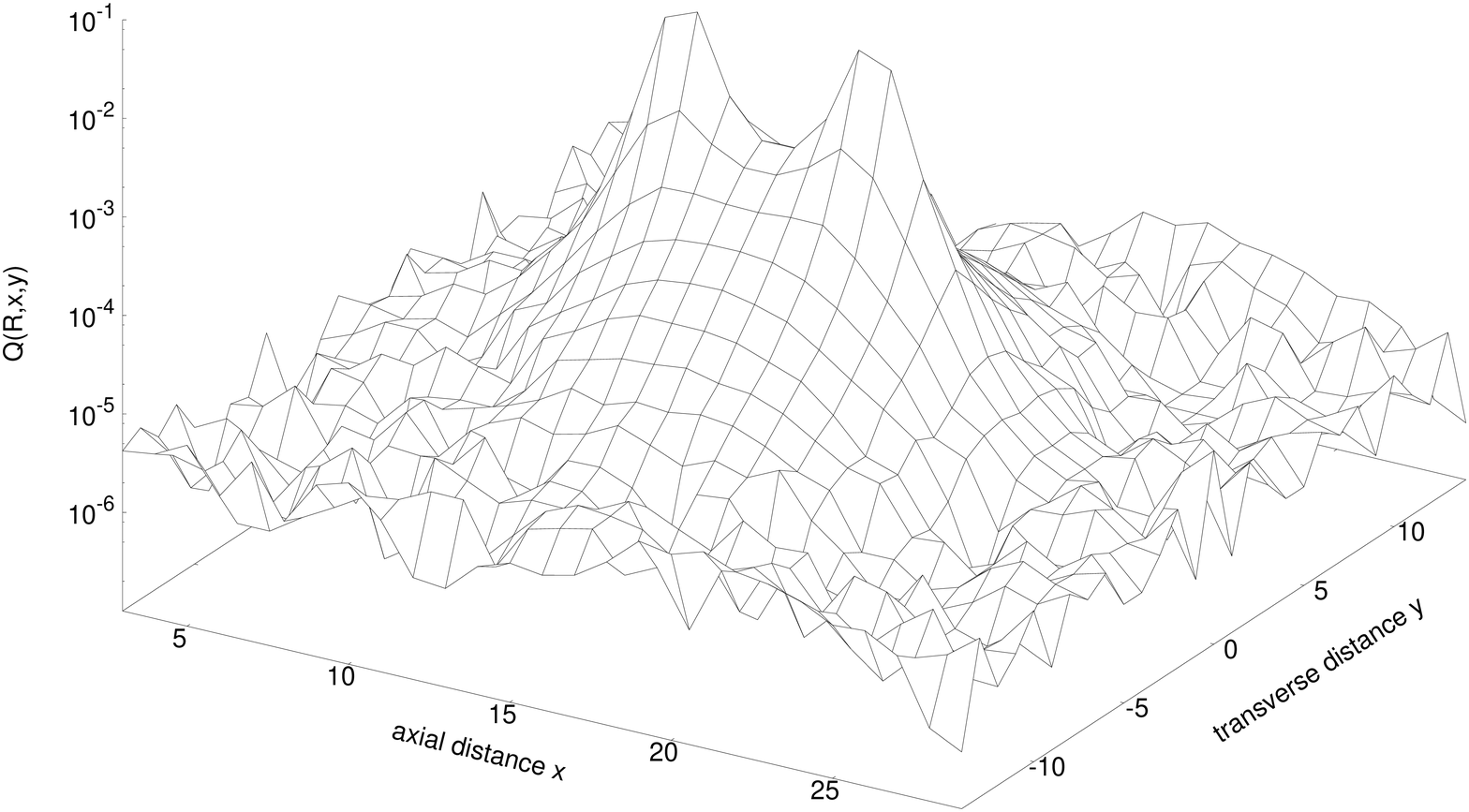}
}
\subfigure[]
{   
 \label{edge}
 \includegraphics[scale=0.135]{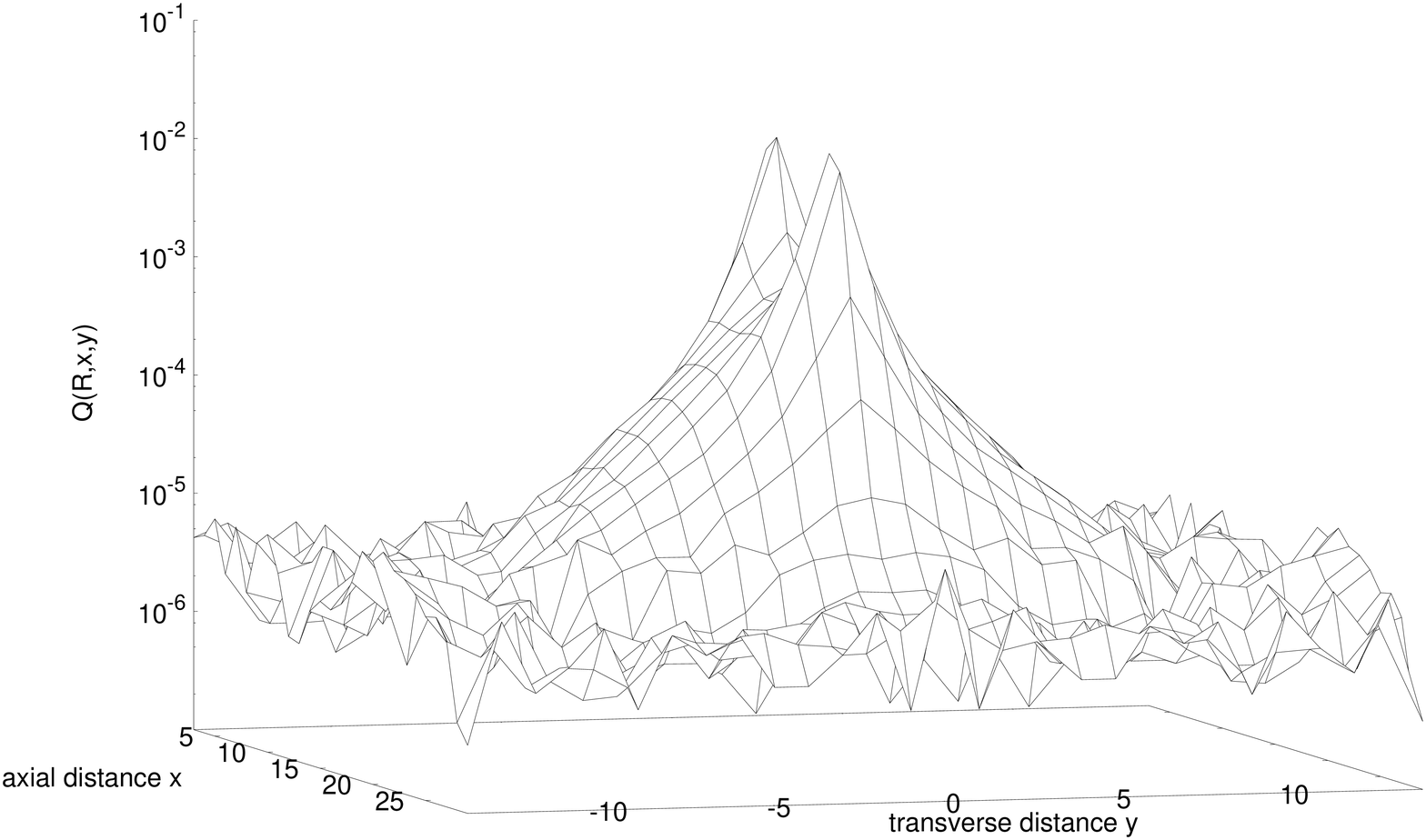}
}
\end{center}
\caption{Two views, from different perspectives, of the Coulomb flux tube at quark-antiquark separation $R=5$.  Note the
logarithmic scale on the $z$-axis.}
\label{tube}
\end{figure}

     Let the quark-antiquark pair lie on the $x$-axis with separation $R$.  We measure, for the $E_x$ component, the $\tr E^2_x$ field at a point $\vp$ which is transverse distance $y$ from the midpoint of the line joining the quarks, subtracting away the vacuum contribution, i.e.
\beq
\langle \Psi_{\qbar q}| \tr E_x^2 (\vp) |\Psi_{\qbar q} \rangle - 
            \langle \Psi_0 | \tr E_x^2 |\Psi_0 \rangle \ .
\eeq
On the lattice, in Coulomb gauge, this observable (see Fig.\ \ref{corr}) is proportional to
\beq
Q(R,y) = { \langle \tr[U_0({\bf 0},0) U_0^\dg({\bf R}_L,0)]  \oh \tr U_P(\vp,0)\rangle \over 
                          \langle \tr[U_0({\bf 0},0) U_0^\dg({\bf R}_L,0)] \rangle }  -  \langle \oh \tr U_P\rangle \ .
\eeq
where $U_P$ denotes a plaquette variable.
The result of a lattice Monte Carlo calculation of $Q(R,y)$ at $\beta=2.5$ and lattice volume $24^4$, shown in Fig.\ \ref{falloff}, is an exponential falloff in the transverse $y$ direction.  This implies flux tube formation.  A profile of the $E^2_x$ distribution is shown in Fig.\ \ref{tube}.  Note the
log scale on the y-axis.

We can compare the half-width of the Coulomb flux tube with that of the minimal energy flux tube, as reported by  Bali et al.\ in 
\cite{Bali:1994de}, at the same coupling $\beta=2.5$ and $R=7$. We find that the half-width of the Coulomb flux tube is smaller by a factor of  approximately 1.7.

\section{A confinement criterion for gauge theories with matter fields}

Suppose we have an SU(N) gauge theory with matter fields in the fundamental representation, e.g.\ QCD.
Wilson loops have perimeter-law falloff asymptotically, Polyakov lines have a non-zero VEV, so what does it mean
to say that such theories (QCD in particular) are confining?  Many people take this to mean ``color confinement,'' meaning that
all the particles in the asymptotic spectrum are color singlets.  We will refer to this as {\it C-confinement}.  However, if that is what one means by
confinement, then it must be recognized that gauge-Higgs theories described by the Brout-Englert-Higgs mechanism, in which there are only Yukawa forces and a complete absence of linearly rising Regge trajectories, are also confining in this sense.  We know this
because of the work of Fradkin and Shenker \cite{Fradkin:1978dv} and Osterwalder, and Seiler \cite{Osterwalder:1977pc}, who showed that there is no transition in coupling-constant space which isolates the Higgs phase from a confinement-like phase, and of Fr\"olich, Morchio, and Strocchi \cite{Frohlich:1981yi}, who explained that
the particles in the spectrum are created by color singlet operators (see also 't Hooft \cite{tHooft:1979yoe}), and who worked out the appropriate perturbation
theory based on these color singlet particle states.

    On the other hand, in a pure gauge theory, there is a very much stronger meaning that one can assign to the word ``confinement,'' which goes well beyond the statement that the asymptotic spectrum consists of massive color neutral particles.  A pure gauge theory is of
course C-confining, with a spectrum of color singlet glueballs, but it {\it also} has the property that the potential energy of a static
quark-antiquark pair rises linearly with separation.  Equivalently, Wilson loops have an area-law falloff asymptotically.  We ask whether there is any way to generalize the area law criterion to gauge theories with string breaking, and matter fields in the fundamental representation of the gauge group.

    The area law criterion can be reformulated, in a pure gauge theory, in the following way.:  Define an initial state\footnote{Indices $abcd$ in this section are color indices in the fundamental representation.}
\beq
          \Psi_{\overline{q} q}(t=0) \equiv \overline{q}^a(\vx) V_0^{ab}(\vx,\vy;A) q^b(\vy) \Psi_0 \ ,
\label{extra1}
\eeq
where the $\overline{q}, q$ operators create an extremely massive static quark-antiquark pair with separation $R=|\vx-\vy|$, $\Psi_0$ is the vacuum state, and
\beq
    V_0(\vx,\vy;A) = P\exp[i \int_\vx^\vy dz^\m A_\m(z)]
\label{Wline}
\eeq
is a Wilson line joining points $\vx,\vy$ at time $t=0$ along a straight-line path.  Let this state evolve in Euclidean time.
Since the quarks are static and the initial state is gauge invariant, the state at Euclidean time $t$ has the form
\beq
 \Psi_{\overline{q} q}(t) = \overline{q}^a(\vx) V_t^{ab}(\vx,\vy;A) q^b(\vy) \Psi_0 \ ,
\label{extra2}
\eeq
where $V_t^{ab}(\vx,\vy;A)$ is a gauge bi-covariant operator which transforms, under a local gauge transformation $g(\vx,t)$, as
\beq
          V_t^{ab}(\vx,\vy;A) \ra V'^{ab}_t(\vx,\vy;A) = g^{ac}(\vx,t) V_t^{cd}(\vx,\vy;A) g^{\dg db}(\vy,t) \ .
\label{bicovariant}
\eeq
As evolution proceeds in Euclidean time, the state  $\Psi_{\overline{q} q}(t)$ evolves, as $t\ra \infty$, to the minimum energy configuration.
Then it is easy to see that the Wilson area law criterion is equivalent to the following property which we will call ``separation of charge confinement,'' or simply {\it S${}_c$-confinement}, defined in the following way:
Let $V(\vx,\vy;A)$ be a gauge bi-covariant operator transforming as in \rf{bicovariant}, and let $E_V(R)$, with $R=|\vx-\vy|$ be the 
energy of the corresponding state
\beq
        \Psi_V \equiv  \overline{q}^a(\vx) V^{ab}(\vx,\vy;A) q^b(\vy) \Psi_0
\label{Vstate}
\eeq
above the vacuum energy $\E_{vac}$.
\textbf{\textit{S${}_c$-confinement}} means that there exists an asymptotically linear function $E_0(R)$, i.e.\
\beq
    \lim_{R\ra \infty} {dE_0 \over dR} = \s >0 \ ,
\eeq
such that
\beq
        E_V(R) \ge E_0(R) 
\label{criterion}
\eeq
for {\it any choice whatever} of bi-covariant $V(\vx,\vy;A)$.   If the ground state saturates this bound then the Wilson loop will have an area law falloff with the coefficient of the area equal to $\s$.

    Our proposal is that the S${}_c$-confinement criterion applies also to gauge theories with matter in the fundamental representation, with the
crucial restriction that the set of $V(\vx,\vy;A)$ depends {\it only} on the gauge field $A_\m(\vx)$, at a fixed time, and not on the matter fields.  If the energy expectation value of states of this kind are also bounded from below by a linear potential, then the theory is S${}_c$-confining as
well as C-confining.  If matter fields were allowed in the definition of $V$, then it would be easy to construct widely separated color singlet 
bound states consisting of a static quark+matter pair, and a static antiquark+matter pair, and in that case $E_V(R)$ would be only weakly
dependent on $R$ at large $R$.  We restrict $V$ to depend only on the gauge field in order to exclude states of that kind. 
     
     S${}_c$-confinement is difficult to verify even numerically, because the set of bi-covariant operators $V(\vx,\vy;A)$ is infinite.  The best we can do at the moment is to pile up examples, and show that they satisfy the S${}_c$-confinement criterion.  On the other hand, if we can find even one case where the S${}_c$-confinement bound is violated, then the system is at most C-confining.  Based on the existence of linear Regge trajectories corresponding to unbroken string states, we conjecture that QCD is S${}_c$-confining.  Likewise we conjecture that a gauge-Higgs theory, e.g.\ for fixed modulus Higgs fields in the fundamental representation with action
\beq
     S = \beta \sum_{plaq} \oh \mbox{Tr}[UUU^\dg U^\dg] 
       + \gamma \sum_{x,\m} \oh \mbox{Tr}[\phi^\dg(x) U_\m(x) 
\phi(x+\widehat{\m})] \ , 
\label{ghiggs}
\eeq
with $\phi$ an SU(2) group-valued field,
has a transition between an S${}_c$-confining phase, which has flux tube formation and string breaking much like QCD, and a Higgs phase.
It must be emphasized that this does not necessarily correspond to a thermodynamics transition.  We know from the work of Fradkin and Shenker \cite{Fradkin:1978dv} and Osterwalder, and Seiler \cite{Osterwalder:1977pc} that the two phases are not completely isolated from one another by a line of non-analyticity in any local gauge-invariant observables.  However there {\it can} be non-analyticities in non-local operators, and this non-locality is implicit in the definition of S${}_c$-confinement.  The question is which observables would be helpful in detecting a transition between S${}_c$-confinement and
C-confinement, if such a transition exists.

    Certainly the Wilson line \rf{Wline} is not helpful; the corresponding $E_V(R)$ grows linearly even in a non-confining abelian theory. We will focus instead on two types of operators.  The first is 
\beq
           V_C^{ab}(x,y) = \G^{\dg ac}(x;A) \G^{cb}(y;A) \ ,
\label{VC}
\eeq
where $\G(x;A)$ is the non-abelian gauge transformation which takes the gauge field to Coulomb gauge.  The reason for this choice is that in an abelian theory, where
\beq
          \G(x;A) = \exp\left[i\int d^3x' ~ A_i(x') \pa_i {1\over 4\pi |x-x'|} \right] \ ,
\eeq
the corresponding state $\Psi_V$ is the minimal energy state containing two static $+/-$ electric charges, and it violates
S${}_c$-confinement (as it should).  In a non-abelian theory, $\Psi_V$ in Coulomb gauge has the deceptively  local appearance
$\Psi_V = \overline{q}^a(\vx)  q^a(\vy) \Psi_0$, and the energy $E_C(R)$ above the vacuum energy is given by $\E_C(R)$ in \rf{Ecoul}.

   The second operator we have looked at introduces a ``pseudo-matter'' field $\vph^a(\vx;A)$ which transforms like a matter field
in the fundamental representation of the gauge group, yet is built entirely from the gauge field, and which, unlike a dynamical 
matter field, has no influence of the probability distribution of the gauge field.  Examples  of such pseudo-matter fields are the
eigenstates $\vph^a_n(\vx)$ of the covariant Laplacian operator, satisfying $(-D_i D_i)^{ab}_{\vx \vy} \vph^b_n(\vy) = \l_n \vph^a_n(\vx)$.
We will choose, as a particular example, the $V$ operator built from the lowest eigenstate  
$V^{ab}_{pm}(\vx,\vy;A) = \vph_1^a(\vx) \vph_1^{\dg b}(\vy)$, and the lattice version of the energy from the logarithmic time derivative is
given by 
\beq
     E_{pm}(R) = -\log \left[ {\langle \{ \vph_1^\dg((\vx,t ) U_0(\vx,t) \vph_1(\vx,t+1) \} \{\vph^\dg_1(\vy,t+1) U_0^\dg(\vy,t) \vph_1(\vy,t) \}\rangle
                              \over \langle \{ \vph_1^\dg((\vx,t ) \vph_1(\vx,t) \} \{ \vph^\dg_1(\vy,t)  \vph_1(\vy,t) \} \rangle } \right] \ .
\label{Epm}
\eeq

\begin{figure}[t!]
\begin{center}
 \includegraphics[scale=0.6]{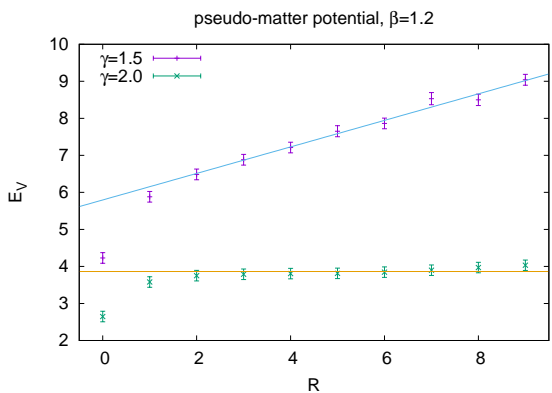}
\caption{The energy $E_{pm}(R)$ defined in \rf{Epm} vs.\ $R$ for pseudo-matter states in the gauge-Higgs model at $\b=1.2$, compared at 
$\g=1.5$ (confining region) and $\g=2.0$ (Higgs region), and computed on a $20^4$ lattice volume. The straight lines are best fits of the data at $R \ge 2$ to a linear rise ($\g=1.5$) and a constant ($\g=2.0$). } 
\end{center}
\label{clap}
\end{figure}
 
    It turns out that both $E_C(R)$ and $E_{pm}(R)$ obey the S${}_c$-confinement bound, in the $\b-\g$ phase diagram of the gauge-Higgs theory, in the small-$\g$ (``confinement-like'') region, and show a transition to C-confinement at large $\g$.  The fact that $E_C(R)$ behaves this way was shown some years ago in ref.\ \cite{Caudy:2007sf}, and is associated with the spontaneous breaking of a remnant global symmetry 
that exists in Coulomb gauge, consisting of gauge transformations which are constant on each time slice.  $E_C(R)$ rises linearly at small
$\g$ in the confinement-like regime, and asymptotes to a constant in the large $\g$ Higgs regime.  The energy expectation value 
$E_{pm}(R)$ also behaves in this way, as seen in Fig.\ \ref{clap}.  In contrast, the energy expectation value $E_{V_0}$ of a state constructed from the Wilson line operator $V_0$ in eq.\ \rf{Wline}, and even states with Wilson line operators constructed from smeared
links, continues to rise linearly in the Higgs regime.  The lesson here is that, while the S${}_c$-confinement property requires that $E_V(R)$
is bounded from below by a linear potential for {\it all} choices of $V$, even a single $V$ which does not respect this bound (and
$E_C(R)$ and $E_{pm}(R)$ in the Higgs region are two such examples) is sufficient
to show that the phase is C-confining but not S${}_c$-confining.

\section{Conclusions}

   It must be emphasized that the Coulomb flux tube is not the usual flux tube of the minimal energy state of a quark-antiquark pair.
It is, rather, the color electric energy distribution of a particular state, of a type first considered by Dirac  \cite{Dirac:1955uv}, corresponding to  $\Psi_V$ in \rf{Vstate} with $V$ in \rf{VC}.  In Coulomb gauge this state simplifies to isolated quark-antiquark operators acting on
the vacuum, but it should be kept in mind that this is just a special form of the more general gauge-invariant expression.  We have
found that the color electric field associated with this state is collimated into a flux tube, which is narrower than the flux tube associated
with the minimal energy quark-antiquark state.  For further details, see \cite{Chung:2017ref}.  The fact that the Coulomb string tension is higher than the asymptotic string tension is no surprise, because the state under consideration is not the minimal energy state.  What {\it is} a little surprising is the fact that the color electric field is collimated in this case.  From lattice
investigations of the ghost propagator in Coulomb gauge one might have expected a power law falloff rather than an exponential falloff
of the electric field away from the line joining the quark and antiquark.  It is not obvious (and it is not clear to us) why the exponential falloff is realized.  One conclusion we must draw from this behavior is that confinement in Coulomb gauge is more subtle than simply a linear potential due to dressed (longitudinal) one-gluon exchange, whose instantaneous part results in the expression \rf{Ecoul}. 

   The Coulomb state is only one of an infinite class of states $\Psi_V$ of the form \rf{Vstate}, based on bi-covariant operators
$V^{ab}(\vx,\vy;A)$.  In this talk we have suggested a new confinement criterion, S${}_c$-confinement, applicable not only to pure gauge theories but also to gauge theories with matter fields in the fundamental representation of the gauge group.  The criterion is simply that the
energy expectation value of all possible states $\Psi_V$ is bounded from below by an asymptotically linear potential.  This generalizes the Wilson area law criterion.  A crucial element in the new criterion is that $V^{ab}(\vx,\vy;A)$ can depend {\it only} on the gauge field, and
not on the matter fields.  We have considered, as two examples, the energy expectation values of the Dirac state and a ``pseudo-matter''
state in a gauge-Higgs theory, and shown that at least these operators show a transition from S${}_c$-confinement to C-confinement along some
line or lines in the plane of coupling constants.  This behavior demonstrates the absence of S${}_c$-confinement in the Higgs-like region, and it supports, although it does not prove, the conjecture that there is S${}_c$-confinement in the remainder of the coupling-constant plane.  For a more
comprehensive discussion, see \cite{Greensite:2017ajx}.

\bibliography{scon}

\end{document}